\documentclass[runningheads]{llncs}
\usepackage[T1]{fontenc}
\newcommand{\ie}{\textit{i}.\textit{e}.}
\usepackage{graphicx}
\usepackage{algorithm}
\usepackage[utf8]{inputenc}     
\usepackage{amsmath,amssymb,amsfonts}
\usepackage{textcomp}
\usepackage{xcolor}
\usepackage{caption}
\usepackage{booktabs}
\usepackage{comment}
\begin{document}
\title{Collaborative Split Federated Learning with Parallel Training and Aggregation}
\titlerunning{Collaborative Split Federated Learning} 
\author{
Yiannis Papageorgiou\inst{1}\orcidID{0009-0009-6783-1320} \and
Yannis Thomas\inst{1}\orcidID{0000-00001-6489-7716} \and
Alexios Filippakopoulos\inst{1}\orcidID{0009-0000-8850-0306} \and
Ramin Khalili\inst{2}\orcidID{0000-0003-2463-7033} \and
Iordanis Koutsopoulos\inst{1}\orcidID{0000-0001-7699-5276}
}
\authorrunning{Yiannis Papageorgiou et al.} 
\institute{
Department of Informatics, Athens University of Economics and Business, Athens, Greece\\
\email{\{gpapageorgiou,thomasi,filippakopoulos,jordan\}@aueb.gr}
\and
Munich Research Center, Huawei Technologies Duesseldorf GmbH, Munich, Germany\\
\email{ramin.khalili@huawei.com}
}
\maketitle
\begin{abstract} 
Federated learning (FL) operates based on model exchanges between the server and the clients, and it suffers from significant client-side computation and communication burden.
Split federated learning (SFL) arises a promising solution by splitting the model into two parts, that are trained sequentially: the clients train the first part of the model (client-side model) and transmit it to the server that trains the second (server-side model). 
Existing SFL schemes though still exhibit long training delays and significant communication overhead, especially when clients of different computing capability participate.

Thus, we propose \emph{Collaborative-Split Federated Learning}~(C-SFL), a novel scheme that splits the model into three parts, namely the model parts trained at the computationally weak clients, the ones trained at the computationally strong clients, and the ones at the server.
Unlike existing works, C-SFL enables parallel training and aggregation of model's parts at the clients and at the server, resulting in reduced training delays and commmunication overhead while improving the model's accuracy.
Experiments verify the multiple gains of C-SFL against the existing schemes.
\end{abstract}
\section{Introduction}
The wide deployment of devices that gather vast amounts of data, along with the stringent data privacy requirements of numerous applications (such as healthcare and natural language processing applications), drives the adoption of distributed learning schemes, like Federated Learning (FL) \cite{mcmahan:2017}.
In FL, clients update their local model for multiple local epochs and send it to a server, which aggregates all local models and transmits the aggregated model back to the clients for the next training round.
This synchronous process is repeated for multiple rounds.

Some of the main challenges in FL include: 1) high computation burden on computationally weak clients and 2) significant communication overhead for uploading/downloading large models via wireless links \cite{zhang:2021},\cite{ye:2023},\cite{one6g:2024}.
In FL, \textit{client heterogeneity}, meaning clients with different computing capability participate, significantly increases the training delay \cite{imteaj:2021}.
The idea of Split Federated Learning (SFL) to split the model, that consists of $V$ sequential layers, into two parts between the clients and the server at a cut layer, tackles FL challenges without compromising accuracy \cite{vepakomma:2018}, \cite{thapa:2022}.
The clients update, perform Forward Propagation (FP) and Backward Propagation (BP), on the first part of the model (client-side model) until the cut layer and transmit it to the server that updates the second (server-side model).
The server aggregates the client-side models per trainind round and the server-side models per local epoch.

However, existing SFL schemes still face long training delays and significant communication overhead.
Firstly, clients wait (remain idle) to receive the gradients from the server to continue the BP process, which becomes increasingly time consuming as the number of clients grows. 
In this regard, the idea of training both the client-side and the server-side models in parallel \cite{han:2021} can effectively reduce the training delay but miserably affects the model's accuracy, since the client-side model is updated using the local loss at the cut layer \cite{nokland:2019}.
Other works also suggest to select individual cut layers per client based on their computation capability/speed, which reduces training delay, but degrades model's accuracy since models with different layers are aggregated at the server\cite{lin:2024}, \cite{shin:2023}. 

And secondly, significant communication overhead incurs, since many models, one per client, are exchanged per round.
The model's accuracy increases as the the number of local epochs per round reduces, meaning that the client-side models are aggregated more frequently, but at the cost of extra communication overhead\cite{lin:2024}.
Thus, a more efficient scheme is needed to reduce training delay and communication overhead while enhancing accuracy within a given time frame.

In this paper, we introduce a novel SFL scheme called Collaborative - Split Federated Learning (C-SFL), which splits the model into three parts at two layers, the collaborative and the cut layer, taking into account client heterogeneity.
Specifically, the first parts until the collaborative layer are updated at the computationally weak clients (weak-side model), the parts between the two layers are updated at the computationally strong clients (aggregator-side model), called local aggregators, and the rest parts at the server. 
Each (weak) client updates its own weak-side model and each local aggregator updates its own weak-side model and the aggregator-side models for its assigned clients. 
Also, each local aggregator aggregates these aggregator-side models, in parallel with the server-side models' update and aggregation per local epoch.

Unlike existing works, the existence of local aggregators improves the model's accuracy, since portion of the client-side model (aggregator-side model) is aggregated per epoch rather than per multiple epochs, without incurring additional computation or communication costs.
C-SFL also reduces significantly the total communication overhead, as each local aggregator transmits one single aggregated model to the server instead of one individual model for each client.

The objective is to find these two layers to split the model into three parts, that reduce the training delay. 
Thus, we write the training delay as a function of these two layers, in order to compute them through exhaustive search.
The contributions of this paper are summarized as follows:
\begin{itemize}
    \item We propose C-SFL, a novel SFL scheme that reduces the training delay and the communication overhead, while enhancing model's accuracy.
    \item We write the total training delay as a function of the collaborative and the cut layers selections. Since we need to find only two parameters, the exhaustive search is $O(V^2)$ where $V$ is the number of the model's layers. 
    \item Experimental results confirm the superiority of our scheme compared to the existing schemes, in terms of training delay and communication overhead.
\end{itemize}
The remainder of this paper is organized as follows. 
Section \ref{sec:relatedwork} introduces the related work and Section \ref{sec:systemdescription} describes the system design.
Section \ref{sec:eval} includes the evaluation of our proposed scheme and Section \ref{sec:conc} concludes the paper.
\section{Related Work \& Background}\label{sec:relatedwork}
\subsection{Related work}
\textbf{Split Learning and Split Federated Learning:} Several Federated Learning (FL) schemes that aim to reduce the training delay have been proposed in the literature, such as client selection \cite{fu:2023}, asynchronous \cite{xu:2023} or model compression \cite{jiang:2022} schemes.
Yet, high computation burden still exists in computationally weak clients, since they need to train the whole model.
Split learning (SL) overcomes this hurdle, by splitting the model into two parts: the client-side model and the server-side model, without necessarily reducing the model's accuracy \cite{vepakomma:2018}.
Chronologically, the first works focus on vanilla SL \cite{vepakomma:2018}, where clients train their client-side model in a sequential manner, resulting in excessive training delay.
Thapa et al. \cite{thapa:2022} integrate FL and SL, creating Split Federated Learning (SFL), to parallelize client-side training and thereby accelerate the model's training.
Thereupon, several SFL schemes attempt to reduce the delay by controlling the model splitting and aggregation at the server.

Samikwa \emph{et al.} \cite{samikwa:2022} propose an SFL scheme, which aims to jointly minimize the training delay and energy consumption in resource constrained IoT devices, through proper cut layer selection.
Similarly, Kim \emph{et al.} \cite{kim:2023} propose a cut layer selection strategy that minimizes a weight cost function incorporating the training delay, the energy consumption and the communication overhead.
The authors in \cite{wu:2023} present a cluster-based SFL scheme utilizing a joint cut layer selection, client clustering, and transmission rate allocation optimization algorithm to minimize the training delay, considering client heterogeneity and network's transmissions rate.
Similar solutions are proposed in \cite{lin:2024efficient}, \cite{wang:2021}.
To further tackle client heterogeneity, other works \cite{shin:2023}, \cite{lin:2024} also propose to select individual cut layers per client, depending on their computing capability.
The main drawback of such strategies is that client-side models with different number of layers are aggregated at the server, thus different client-side models are updated per client, which reduces the delay but degrades the accuracy.
A method to increase the accuracy is to reduce the number of local epochs per round, meaning that the client-side models are aggregated more frequently, but at the cost of extra communication overhead\cite{lin:2024}.

\textbf{Parallel execution in SFL:}  Some research efforts consider the presence of helpers, namely computationally strong clients assist computationally weak clients to perform the client-side training.
The work in \cite{wang:2023} jointly finds the cut layer and the client-helper assignments considering clients' computing capability.
In \cite{tirana:2024} the authors propose SFL \& Helpers to solve the same problem, considering also the memory capacity of the clients, with the objective to minimize delay.
They extend their work in \cite{tirana:2024mp} by assigning multiple helpers per client.

Another line of works \cite{han:2021},\cite{Seungeun:2022} show that the delay can still be reduced by training in parallel the client-side and server-side models, rather than sequentially.
Specifically, in LocSplitFed \cite{han:2021}, the clients use an additional layer on top of the client-side model that enables them to compute the loss, when the forward propagation until the cut layer (local loss) is completed.
This feature enables clients to perform backward propagation faster using the local loss gradients, rather than waiting for the gradients from the server, in parallel with the server-side model's update.
Yet it leads to accuracy reduction \cite{nokland:2019}, which is an important trade-off.
Another work, FedSplitX \cite{shin:2023} defines individual cut layer per client and allows them to update their client-side model using the local loss at their individual cut layer.
FedSplitX accelerates training with client heterogeneity and results in further accuracy reduction, since models of different layers are aggregated.

In contrast, our approach, namely Collaborative-Split Federated Learning (C-SFL), aims to reduce the training delay and communication overhead, while enhancing accuracy and thus outperforming existing works.
C-SFL splits the model into three parts at two layers, the collaborative and the cut layer, taking into account client heterogeneity.
Specifically, the first parts until the collaborative layer are trained at the computationally weak clients (weak-side model), the parts between the two layers are trained at the computationally strong clients (aggregator-side model), called local aggregators, and the rest parts at the server.
This key design principle of C-SFL enables the parallel training and aggregation of the aggregator-side models alongside the server-side models per local epoch.

C-SFL improves the model’s accuracy, since portion of the client-side model (aggregator-side model) is aggregated per epoch rather than per multiple epochs (round), without incurring additional computation or communication costs
and and is different from current works as shown in Table \ref{table:comparison}.
Additionally, C-SFL reduces significantly the total communication overhead, as each local aggregator transmits one single aggregated model of its assigned clients to the server, instead of transmitting one individual model for each assigned (weak) client.
\begin{table}
 \caption{Comparison of SFL, SFL \& Helpers, LocSplitFed, and Our scheme.}\label{table:comparison}
 \begin{tabular}{|l|l|l|l|l|}
\hline 
Learning Scheme & SFL & SFL \& Helpers&LocSplitFed&\textbf{Our}\\
\hline
 Parallel client-side models' training\cite{thapa:2022},\cite{samikwa:2022} & Yes & Yes&Yes&\textbf{Yes}\\
 \hline
 Client-side training with assistance\cite{tirana:2024},\cite{wang:2023} & No & Yes&No&\textbf{Yes}\\
  \hline
 Parallel client-side and server-side training\cite{han:2021},\cite{Seungeun:2022} & No & No &Yes&\textbf{Yes}\\
  \hline
\textbf{Parallel aggregator-side} & No&No&No&\textbf{Yes}\\
\textbf{and server-side aggregation} & & & & \\
 \hline
\end{tabular}
 \end{table}
\subsection{Background}
The global Deep Neural Network (DNN) model, denoted as 
$W$, consists of $V$ sequential layers and is trained across $N$ clients, each with its own local data.
The objective is to find the optimal global model $W^*$ that minimizes the global loss function: $F(W)=\frac{1}{N} \sum_{i=n}^{N} F_n(W)$, where $ F_n(W)$ represents the global loss computed by model $W$ using the local data of client $n$.

\textbf{Federated Learning (FL):}
solves the above problem via the following iterative process for multiple rounds: Each client $n$ trains its own local model $w_n$ for a certain number of epochs $E$ during one training round $t$, and then transmits it to the server.
The server aggregates the local models, computes the global model $W(t+1)=\frac{1}{N} \sum_{i=1}^{N} W_i(t)$ along with its training loss (using the specified global loss function), and then transmits the updated global model $W(t+1)$ back to the clients for the next round.
Yet, FL suffers from significant delays when training large models, especially with heterogeneous clients, and incurs high communication overhead from model exchanges.

\textbf{Split Federated Learning (SFL)} splits the model at a \textit{cut layer} $v$ into two portions: the client-side model $W^c$, which includes layers \{1, 2, ..., $v$\} and is trained on the client(s), and the server-side model $W^s$, which includes layers \{$v$, ..., $V$\} and is trained on the server.
During each epoch within a round, the clients perform forward propagation (FP) on their portion of the model $w^c_n$, and they compute the cut layer's activations which they then transmit to the server.
The server continues the FP and the backward propagation (BP) process on its portion of the model (server-side models), and it transmits back the respective cut layer's gradients, for clients to complete the BP process and thus, update their client-side models.
By the end of each \textit{round} (multiple epochs) the clients transmit to the server their client-side models, the server aggregates these models into the global client-side model $W^c(t+1)=\frac{1}{N} \sum_{n=1}^{N} W_n^c(t)$, that it transmits back to clients for the next round.
Yet, existing SFL schemes still suffer from long training delays caused, especially with heterogeneous clients, as well as from significant communication overhead during model's exchange.
\section{Collaborative Split Federated Learning}\label{sec:systemdescription}
\begin{figure}[tb]
\hspace{-0.00cm}\includegraphics[angle=0,trim={0cm 0cm 0cm 0cm},clip,scale=0.5]{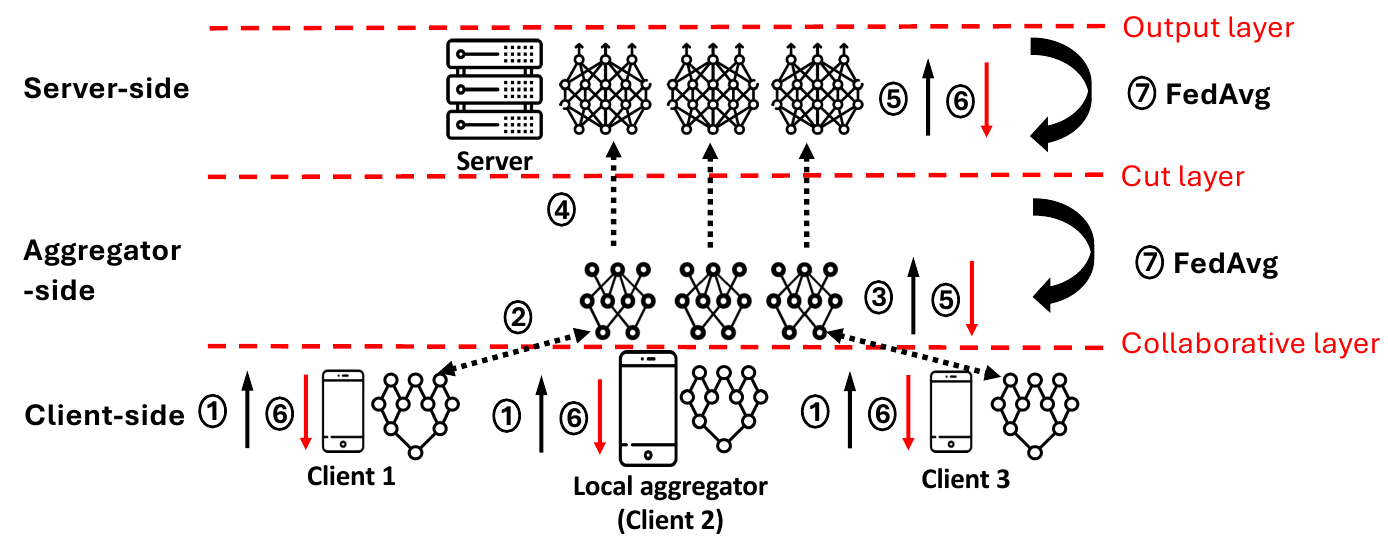}
\caption{Training workflow for three clients, with client 2 selected as the local aggregator. Clients perform FP on their weak-side models, and they send activations to the aggregator (steps 1 \& 2), which continues FP until the cut layer (step 3).
The aggregator transmits the activations to the server (step 4) and then the BP process on the aggregator-side (step 5) and weak-side models (step 6), using the local loss at the cut layer, is initiated. In parallel, the server executes FP and BP (steps 5 \& 6).
Finally, the server-side and aggregator-side models are aggregated also in parallel (step 7).}
\label{fig:workflow}
\end{figure}
\subsection{System model}
The network comprises a set of clients $\mathcal{N}$ and one server which are connected through (wireless) links. 
The computing speed of client $n$ is denoted with $p_n$ (in Flops/sec), and the computing speed of the server is denoted with $p_s$.
The transmission rate of each link connecting two entities (clients or the server) $i$ and $j$ is denoted with $r_{i,j}$ (in bps).
Each computationally weak client $n$, namely a client with small $p_n$, collaborates with one unique computationally strong client $k$ by offloading a portion of its computation workload (layers) to the strong client $k$, enabling it to update its own client-side model $w^c_n$ using its own data of $B$ batches.
Specifically, the client-side ($W^c$) model is splitted at the \textit{collaborative layer $h$} into two portions: the weak-side model ($W^w$), which includes layers \{1, 2, ..., $h$\} and is trained on the weak client $n$, and the aggregator-side model ($W^a$), which includes layers \{$h$, ..., $v$\} and is trained on the strong client $k$. 
These strong clients are referred to as \textit{local aggregators}, and each weak client $n$ is assigned to a unique local aggregator $k$.
This assignment is represented by the binary variable $x_{n,k}$.
The fraction of clients functioning as local aggregators is denoted by $\lambda$, while the remaining $(1 - \lambda)N$ (weak) clients are each assigned to a single local aggregator.
Also, the set $S_k$ includes the assigned clients to the local aggregator $k$ and the set of local aggregators is denoted with $\mathcal{K}$.
Each local aggregator leverages the existing aggregator-side models (part of the client-side), of its assigned clients $S_k$, by aggregating them: $W^a_k(e+1)=\frac{1}{|S_k|} \sum_{n \in S_k}^{} W_n^c(t)$, as depicted in Figure \ref{fig:workflow} per epoch $e$.
When $\lambda=1$, the C-SFL scheme is reduced to the standard SFL scheme. 
The system variables are presented in Table \ref{table:notation}.
\subsection{System design}

\textbf{Loss functions:} We consider two loss functions. 
Except the conventional \textit{global loss} function $F(w)$ that is computed given the output layer's activations, we also consider the \textit{local loss} function that is computed given the cut layer's activations.
The \textit{local loss} function utilizes a multilayer perceptron positioned above the aggregator-side model.
This auxiliary network makes predictions using the cut layer's activations and computes the actual local loss at the cut layer.
Thus, clients can update their client-side model (through BP) without waiting for the server-side models' update, using a local loss function\cite{han:2021}.
Its utilization requires negligible computation workload.
 \begin{table}[tb]
 \caption{Notation Table}\label{table:notation}
 \begin{tabular}{|l|l|}
\hline
$a_j$ & Weights of layer $j$ (bits) \\
\hline
$D_m$ & Training delay of phase $m$\\
\hline
$f_j$ & Computation workload of layer $j$ (Flops)\\
\hline
$v$, $h$ & Cut, Collaborative layer\\
\hline
$\mathcal{N}, \mathcal{K}$ & Set of clients, aggregators \\
\hline
$p_n, p_k$ & Computation speed of client $n$, local aggregator $k$ (Flops/sec)\\
\hline
$p_s$ & Computation speed of server (Flops/sec)\\
\hline
$r_{i,j}$ & Transmission rate of link between clients $i$ and $j$ (bps)\\
\hline
$w^w_{n,e}$ & Weak-side model of client $n$ at epoch $e$\\
\hline
$w^a_{n,e}$ & Aggregator-side model of client $n$ at epoch $e$\\
\hline
$w^s_{n,e}$ & Server-side model of client $n$ at epoch $e$\\
\hline
$x_{n,k}$ & Assignment of client $n$ with local aggregator $k$ (binary variable)\\
\hline
$\sum_{n \in \mathcal{N}} x_{n,k}$ & Number of clients assigned to aggregator $k$ \\
\hline
\end{tabular}
 \end{table}
 
\textbf{Training workflow}
The training workflow in C-SFL is centrally orchestrated by the server, usually an edge-server that is connected with the clients.
Specifically, the server decides and informs the clients for the selected collaborative $h$ and cut $v$ layers.
The network bootstraps by initializing the weak-side ($W^w$), the aggregator-side ($W^a$) and the server-side ($W^s$) models randomly.
The training workflow is divided into four phases.
Phase 0 starts with the clients downloading the weak-side model $W^w$, and the aggregators downloading the aggregator-side $W^a$ model from the server.

During phase 1, each client $n$ performs FP on its $w^w_n$ until the (collaborative) layer $h$, it obtains the corresponding activations and then transmits these activations (see steps 1-2 in Fig.\ref{fig:workflow}), to its assigned local aggregator $x_{n}$. 
Each local aggregator $k$ performs FP on its own weak-side model $w^w_k$ and aggregator-side model $w^a_k$ and computes the local loss. 
It also performs FP on the aggregator-side models that correspond to the clients assigned to it,  
until the cut layer $v$ (steps 3 \& 4 in Fig.\ref{fig:workflow}) and transmits  the corresponding activations to the server. 

During phase 2, each local aggregator $k$ computes the local loss gradients and then performs BP on the aggregator-side models that correspond to the clients assigned to it.
Then, it transmits the collaborative layer's $h$ gradients back its assigned clients for them to complete the BP process and thus update their respective weak-side model.

In parallel, upon receiving the cut layer's activations from the local aggregator(s), the server updates the server-side models for all clients and computes the respective global loss at the output layer to perform BP. 
At the end of the current epoch, the server aggregates the server-side models for all clients to compute the global server-side model $W^s$ for the next epoch, while in parallel each local aggregator $k$ aggregates the aggregator-side models for its assigned clients ($W^a_k(e+1)=\frac{1}{|S_k|} \sum_{\forall n \in S_k}^{} W_n^a(e)$), as depicted in Figure \ref{fig:workflow} (step 7).
After $E$ epochs (one round), during phase 3, the local aggregators transmit their aggregator-side model $w_k^a$ to the server and each client $n$ transmits its weak-side model $w_n^w$ to the server.
The server computes the global aggregator-side model $W^a$ and global weak-side model $W^w$ using the FedAvg algorithm \cite{collins:2022}.
And then it transmits the updated weak-side model back to the clients and the updated aggregator-side model to the local aggregators for the next round.
\subsection{Training delay analysis}
In this section, the training delay analysis for each phase of the C-SFL is presented.
The training epoch index $e$ is omitted for notational simplicity.
The aggregation of client-side models using the FedAvg algorithm has low computational complexity, leading to negligible delay.
During phase 0, the server broadcasts the initial weak-side model $W^w(t)$ to the clients and the aggregator-side model $W^a(t)$ to the aggregators.
Let $a_j$ denote the number of bits representing the weights of layer $j$, and let $f_j$ denote the corresponding computational workload (Flops).
The resulting delay is the one that corresponds to the slowest download delay between the two models:
\begin{equation}\label{eqn:d0}
D_0=\max \left\{ \max_{n \in \mathcal{N}} \{
        \frac{\sum_{j \in \{1..h\}^{}}a_j}{r_{s,n}} 
   \}, \quad
    \max_{k \in \mathcal{K}} \{
         \frac{\sum_{j \in \{h..v\}}a_j}{r_{s,k}}\}  \right\}
\end{equation}
The delays incurred during phase 1 include:

Weak-side model's FP delay:
Each client performs FP on its weak-side model, using its local data.
This delay corresponds to the fraction of the weak-side model's computational workload divided by the client's computation speed $p_n$.

Collaborative layer's activations transmissions delay:
After completing the weak-side model's FP, the clients transmit to their assigned local aggregator the collaborative layer's activations.

Aggregator-side models' FP delay:
The local aggregators execute FP on the aggregator-side models for their assigned clients.

Cut layer's activations transmissions delay:
After completing the aggregator-side models' FP, the local aggregators transmit to the server the corresponding cut layer's activations.

Consequently, the delay of phase 1 equals with:
\begin{equation}\label{eqn:d1}
D_1=\max\limits_{n \in \mathcal{N}, k \in \mathcal{K}} \{
        \frac{\sum\limits_{j \in \{1 \ldots h\}} f_j}{p_n}  + \frac{a_h}{r_{n, x_{n}}} 
        + \frac{\sum\limits_{j \in \{h \ldots v\}} f_j \cdot \sum\limits_{n \in \mathcal{N}} x_{k,n}  }{p_n}  
        + \frac{\sum\limits_{n \in \mathcal{N}} x_{k,n} \cdot a_v}{r_{k,s}} \}
\end{equation}
The delays incurred during phase 2 include:

Server-side models' update delay:
This step involves the execution of the server-side models' FP and BP with the collected activations from all clients through the local aggregators.

Aggregator-side model's BP delay:
The local aggregators execute FP and BP on the aggregator-side models with the received collaborative layer's activations from their assigned clients.

Gradients' transmissions delay: 
Upon the completion of the BP on the aggregator-side model, the local aggregator $k$ sends the  corresponding collaborative layer's gradients to its assigned clients.

Weak-side model's BP delay:
The (weak) clients execute BP on their weak-side model after receiving the collaborative layer's gradients.
The delay of phase 2 is the maximum between the server-side models' update delay and the sum of delays of the rest processes:
\begin{align}\label{eqn:d2}
D_2 = \max \bigg\{ & \frac{2 \cdot N \cdot \sum\limits_{j \in \{v \ldots V\}} f_j}{p_s}, \nonumber \\
& \max\limits_{n \in \mathcal{N}, k \in \mathcal{K}} \{
\frac{\sum\limits_{j \in \{h \ldots v\}} f_j \cdot \sum\limits_{n \in N} x_{k,n}}{p_k} 
+ \frac{a_h}{r_{x_n, n}}   
+ \frac{\sum\limits_{j \in \{1 \ldots h\}} f_j}{p_n} 
\} \bigg\}
\end{align}
Throughout phase 3, each local aggregator $k$ uploads its aggregator-side models to the server, while clients upload their weak-side models also to the server. The total delay of phase 3 equals with:
\begin{eqnarray}\label{eqn:d3}
D_3= \max \left\{
    \max_{n \in \mathcal{N}} \{
        \frac{\sum_{j \in \{1 \ldots h\}}a_j}{r_{s,n}} 
   \}, 
    \max_{k \in \mathcal{K}} \{
         \frac{\sum_{j \in \{h \ldots v\}}a_j}{r_{s,k}}
        \} \right\}
\end{eqnarray}
Considering the delay components in (\ref{eqn:d0}), (\ref{eqn:d1}), (\ref{eqn:d2}), and (\ref{eqn:d3}), the round delay is:
\begin{equation}\label{eqn:dtotal}
D_{round}= D_0 + E \cdot B \cdot (D_1+D_2) + D_3
\end{equation}

The selection of collaborative and cut layers influences the amount of computational workload of clients, local aggregators, and the server, thereby affecting the training delay.
This selections also impacts the communication overhead, including the amount of bits transmitted and the size of models transmitted for aggregation to both the server and the local aggregators (see Table \ref{tab:computation_table}).

For each pair of $(h, v)$, we find the delay $D_{round}$.
Then, the optimal pair $(h^*,v^*)$ is the one for which $D_{round}$ is the smallest over all possible valid combinations.
In this work, we consider DNNs with a relatively small number of layers $V$.
For $V$ layers, there are $V-1$ possible solutions for the collaborative layer $h$, \ie after layers $1, 2, \ldots V-2$.
Say that $h=j$, there are $(V-1) - (j+1) +1$ solutions for the cut layer $v$, because it has to be after the $h$ layer and before at most be the $V-1$ layer.
Thus, the number of possible valid combinations is:
$\sum_{h=2}^{V-1} \sum_{v=h+1}^{V-1} 1$ which is $O(V^2)$.
\begin{table}[t]
\centering
\caption{Total communication overhead (bits transmitted) during one round.}
\label{tab:computation_table}
\begin{tabular}{p{0.18\columnwidth} p{0.8\columnwidth} p{0.18\columnwidth}}
\toprule
\textbf{Schemes} & \textbf{Communication Overhead } \\ 
\midrule
SplitFed & $2(a_v  B +  \cdot \sum\limits_{j \in \{1 \ldots v\}} a_j)N$  \\ 
\midrule
LocSplitFed  & $(a_v  B + 2 \cdot \sum\limits_{j \in \{1 \ldots v\}} a_j)N$ \\ 
\midrule
C-SFL  & $2(a_h B +\sum\limits_{j \in \{1 \ldots j\}} a_j) (1-\lambda)N + (2\sum\limits_{j \in \{h \ldots v\}} a_j) \lambda N + (a_vB)N$  \\ 
\bottomrule
\end{tabular}
\end{table}
\section{Performance evaluation}\label{sec:eval}
\subsection{Methodology}
We validate our algorithm on MNIST 
FMNIST 
and CIFAR-10 
datasets. 
For MNIST and FMNIST, a CNN is utilized that consists of 5 convolutional layers and 3 fully connected layers as in AlexNet.
The number of model parameters is 3,868,170.
For CIFAR-10, the VGG-11 model is utilized with 9,231,114 parameters.
The training set of each dataset is distributed to the $N=100$ clients for training and the original test set of each dataset is used to assess the accuracy of the global model.
The fraction of clients functioning as local aggregators equals to $\lambda=0.1$, thus the number of local aggregators is $\lambda N=10$.
Each local aggregator (e.g. a mobile device) is assigned the same number of (weak) clients.
Also, their computing speed $p_k$ equals to 16 Ghz (8 cores of 2.2 Ghz each), while the respective computing speed $p_n$ of the rest (weak) clients, such as IoT devices (e.g. Raspberry Pis), equals with 2 Ghz (2 cores of 1 Ghz each).
We denote as $\gamma = p_k/p_n$ the heterogeneity ratio, in terms of computing capability, which indicates the ratio of the computing speed of the local aggregator and the (weak) client.
A higher $\gamma$ indicates greater heterogeneity among clients within the network.
The server's computing capability $p_s$ equals with 100 Ghz (40 cores of 2.5 Ghz each).
We also set the transmission rate to $R= 2$ Mbps for all network links.
We experiment on two different data distribution setups, the IID and the non-IID one.
In the first setup, the data samples from each class are evenly distributed among the $N$ clients, whereas in the second setup, they are not distributed equally \cite{mcmahan:2017}.
The learning rate is 0.0001 and the number of data batches $B$ is set to 36 for the MNIST and FMNIST datasets, while for the CIFAR-10 is set to 8.
Also, clients perform $E=3$ local epochs per round.
\subsection{Results}
\begin{figure}
\begin{tabular}{c c c}
\hspace{-0.41cm}\includegraphics[trim={0cm 0cm 0cm 0cm},clip,scale=0.31]{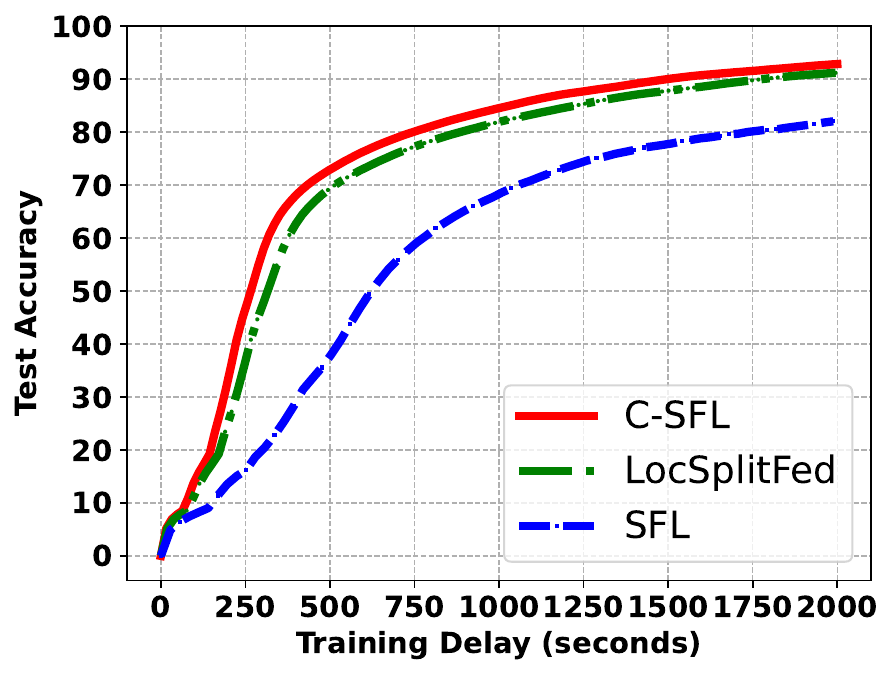} & \hspace{-0.05cm}\includegraphics[trim={2cm 0cm 0cm 0cm},clip,scale=0.31]{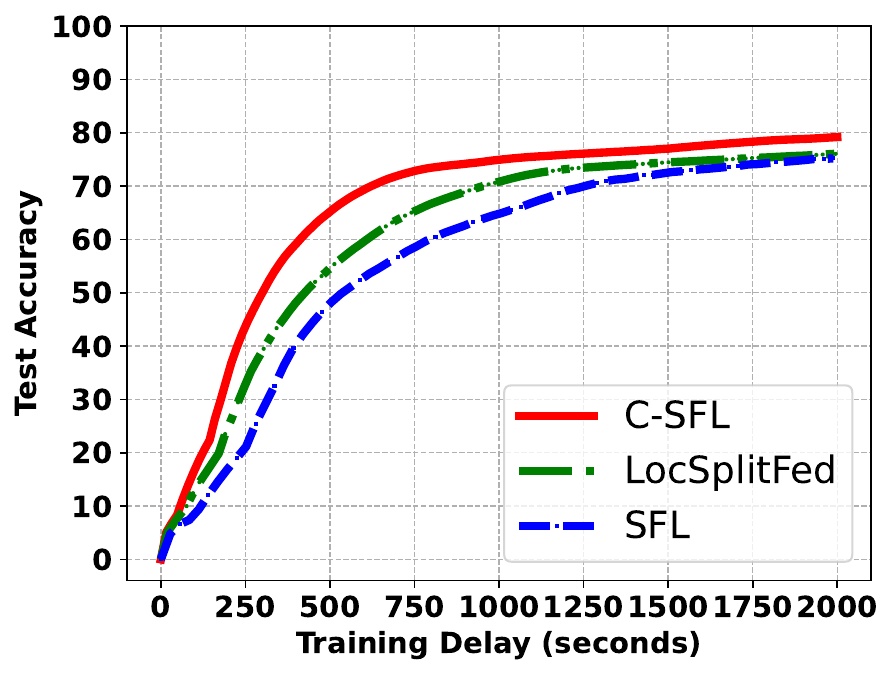} & 
\hspace{-0.05cm}\includegraphics[trim={2cm 0cm 0cm 0cm},clip,scale=0.31]{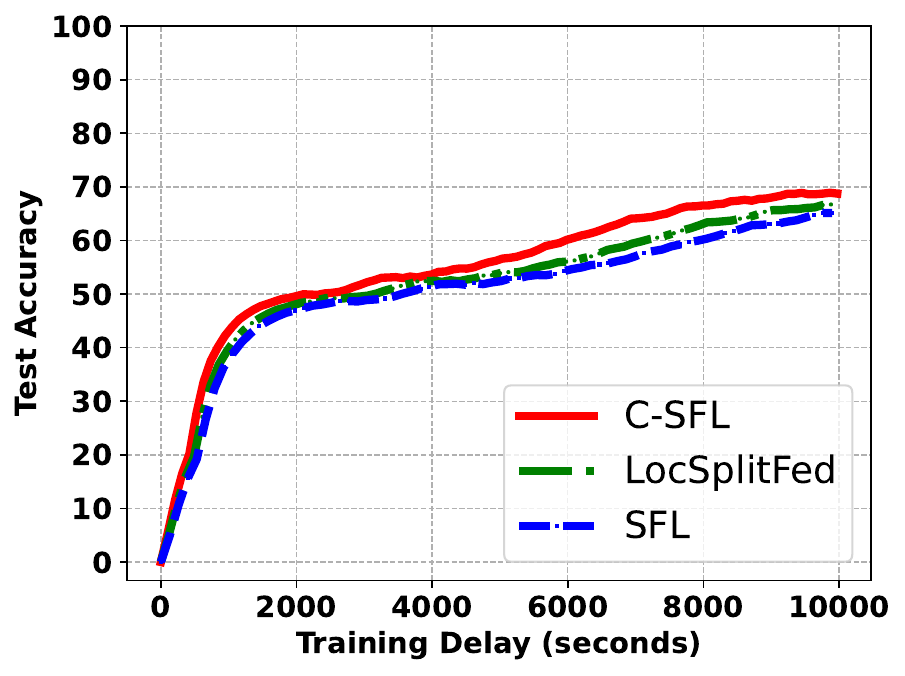} \\
\hspace{-0.01cm} \footnotesize (a) MNIST  & \hspace{-1.cm} \footnotesize (b) FMNIST & \hspace{-1.cm} \footnotesize (c) CIFAR-10\\
\end{tabular}
\caption{Test accuracy versus training delay.}
\label{fig:exp1delay}
\vspace{-0.3cm}
\end{figure}

\begin{table}[t]
\caption{Accuracy for each scheme and datasets with IID and non-IID data.}
\begin{center}
\begin{tabular}{c|c c|c c|c c}
\multicolumn{1}{c|}{} & \multicolumn{2}{c|}{\textbf{\quad MNIST \quad}} & \multicolumn{2}{c|}{\textbf{\quad FMNIST \quad}} & \multicolumn{2}{c}{\textbf{\quad CIFAR-10 \quad}}  \\
\textbf{Scheme} & \quad IID \quad & \quad non-IID \quad & \quad IID \quad & \quad non-IID \quad & \quad IID \quad & \quad non-IID \quad \\
\hline
SFL & 83.06\% & 82.1\% & 79.8\% & 74.23\% & 65.7\%  & 65\%  \\
\hline
LocSplitFed & 91.18\% & 90.1\% & 81\% & 77.27\% & 66.71\% & 65.9\%  \\
\hline
C-SFL & \textbf{92.83\%} & \textbf{91.5\%} & \textbf{83.2\%} & \textbf{78.57\%} & \textbf{68.91\%} & \textbf{68\%} \\
\end{tabular}
\label{tab:totalresults}
\end{center}
\end{table}
We validate the effectiveness of our proposed C-SFL against the following two schemes:
\begin{itemize}
    \item \textbf{SFL:} The global model $W$ is split at the cut layer $v$ into the client-side $W^c$ and server-side $W^s$ models.
Both of them are trained in a sequential manner, and the clients wait for the gradients from the server to perform BP. The publicly available code provided by the authors is reused \cite{thapa:2022}.
 \item  \textbf{LocSplitFed:} The $W^c$ and $W^s$ are trained in parallel, thus the clients perform BP using the local loss at the cut layer $v$, instead of waiting for the gradients from the server.
\end{itemize}
\textbf{Test accuracy versus training delay: }In Fig.\ref{fig:exp1delay} we plot the achieved accuracy of each scheme as a function of the training delay.
It can be seen that our scheme achieves higher accuracy, compared to the other schemes, improving it up to 3\%-4\% on the FMNIST dataset during 2000 seconds of training.
The same holds for the MNIST and CIFAR-10 datasets, as depicted in Table \ref{tab:totalresults}.
The selection of collaborative $h$ and cut $v$ layers, depending on the clients' computing speed, reduces the training delay per round. Also, the key design principle of our scheme to define the aggregator-side model (from  $h$-th to $v$-th layer) and aggregate it per epoch, increases the per round accuracy compared to the LocSplitFed.

\begin{figure}
\begin{tabular}{c c c}
\hspace{-0.41cm}\includegraphics[trim={0cm 0cm 0cm 0cm},clip,scale=0.31]{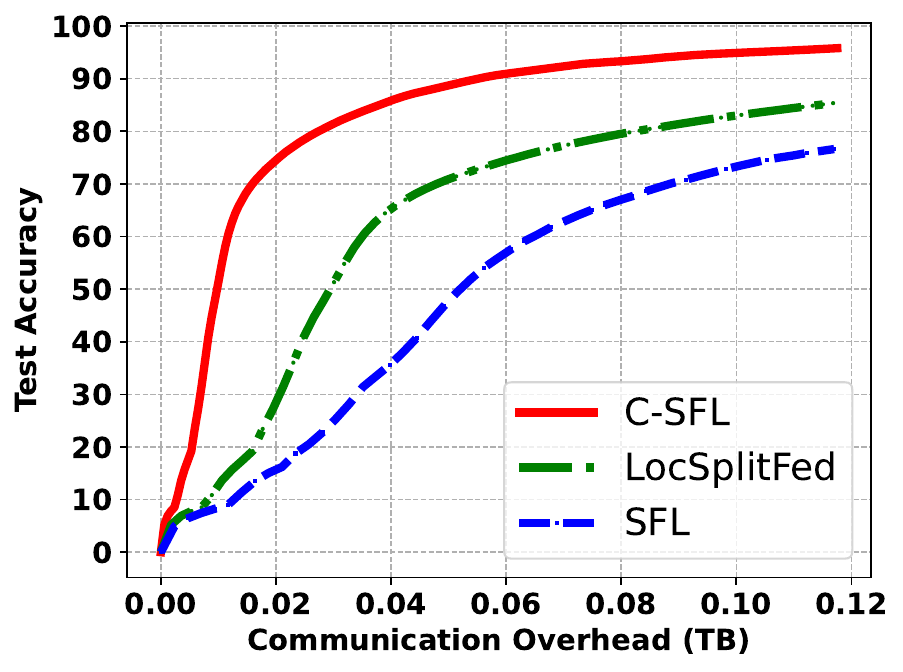} & \hspace{-0.05cm}\includegraphics[trim={2cm 0cm 0cm 0cm},clip,scale=0.31]{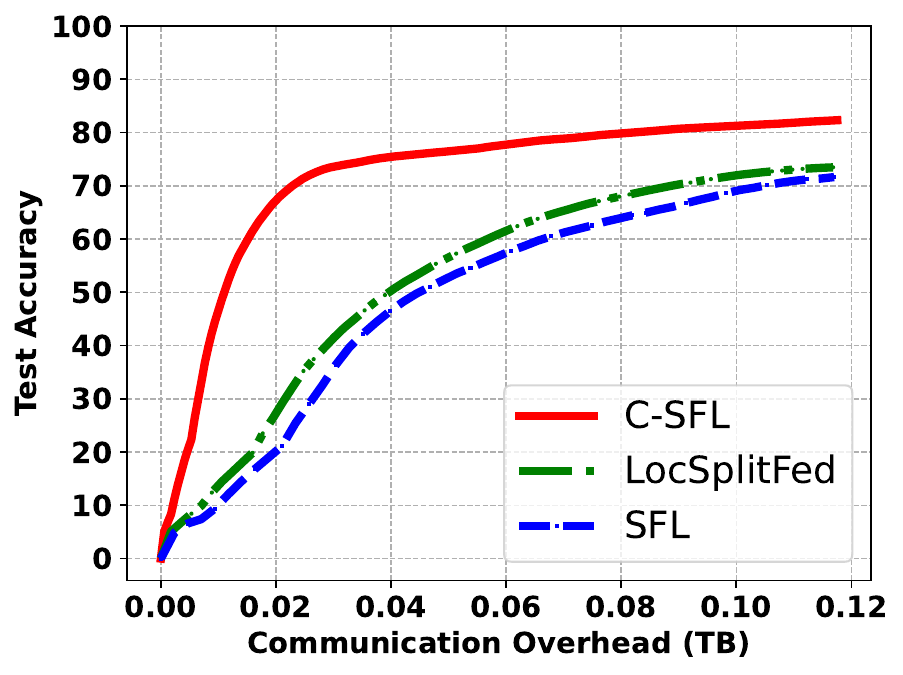} & 
\hspace{-0.05cm}\includegraphics[trim={2cm 0cm 0cm 0cm},clip,scale=0.31]{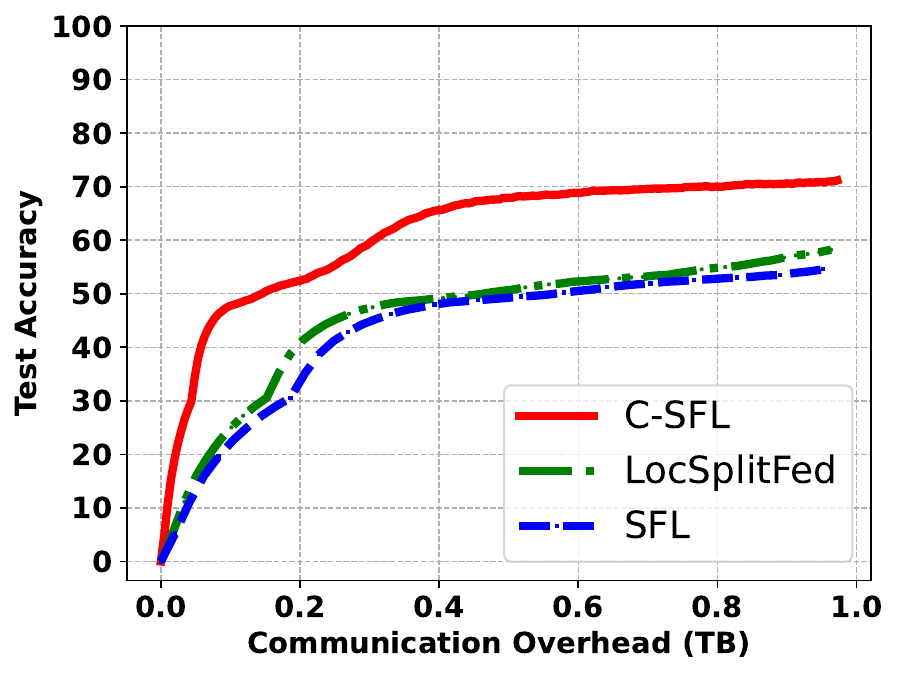} \\
\hspace{-0.01cm} \footnotesize (a) MNIST  & \hspace{-1.cm} \footnotesize (b) FMNIST & \hspace{-1.cm} \footnotesize (c) CIFAR-10\\
\end{tabular}
\caption{Test accuracy versus communication overhead.}
\label{fig:exp1overhead}
\vspace{-0.3cm}
\end{figure}

\textbf{Test accuracy versus communication overhead:} Moreover, our C-SFL scheme significantly reduces the total communication overhead required to achieve the target accuracy, as illustrated in Fig. \ref{fig:exp1overhead}.
Specifically, our scheme achieves approximately 69\% accuracy with a communication overhead of 0.60 TB, whereas other schemes achieve only around 57\%–60\% accuracy with the same communication overhead during the training on the CIFAR-10.
It also achieves nearly 90\% accuracy during training on the MNIST dataset with just 0.06 TB of communication overhead, whereas LocSplitFed reaches 77\% and SFL almost 59\% with the same overhead.
The takeaway message from these experiments is that our proposed scheme outperforms the other schemes both in terms of delay and communication overhead.
\begin{table}[b]
 \centering
 \caption{Collaborative and cut layer selections ($h, v$) of each scheme for various transmission rates $R$ and heterogeneity ratios $\gamma$.}\label{tab:cut_layers}
 \begin{tabular}{|l|l|l|l|}
\hline 
Schemes & SFL&LocSplitFed&\textbf{C-SFL($\lambda=0.1$)}\\
\hline
 ($\gamma=8.5$, R=2)  &  $v=5$& $v=5$&$h=3,v=5$\\
  \hline
 ($\gamma=1$,R=2)  &  $v=5$& $v=5$&$h=3,v=5$\\
  \hline
 ($\gamma=8.5$,R=10)  &  $v=5$ & $v=5$&$h=3,v=4$\\
  \hline
($\gamma=1$,R=10) & $v=5$& $v=5$&$h=4,v=5$\\
 \hline
\end{tabular}
 \end{table}

\begin{figure}
\begin{tabular}{c c c}
\hspace{-0.41cm}\includegraphics[trim={0cm 0cm 0cm 0cm},clip,scale=0.31]{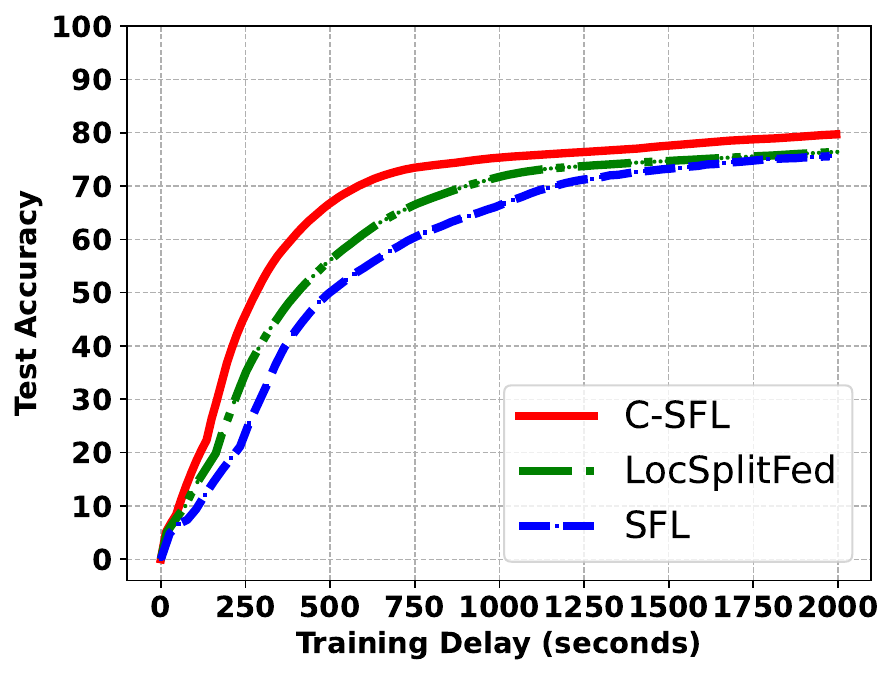} & \hspace{-0.05cm}\includegraphics[trim={2cm 0cm 0cm 0cm},clip,scale=0.31]{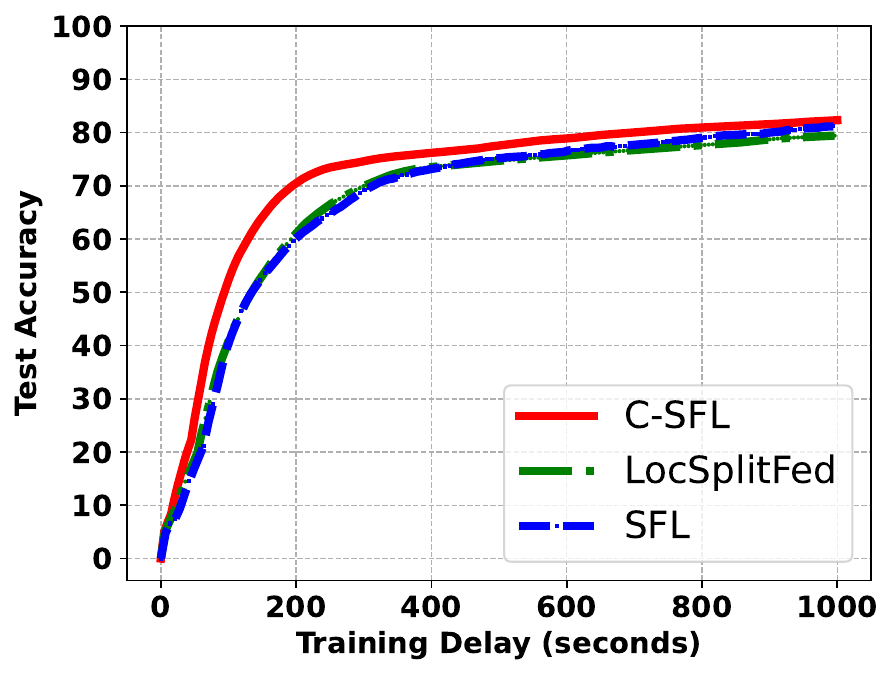} & \hspace{-0.05cm}\includegraphics[trim={2cm 0cm 0cm 0cm},clip,scale=0.31]{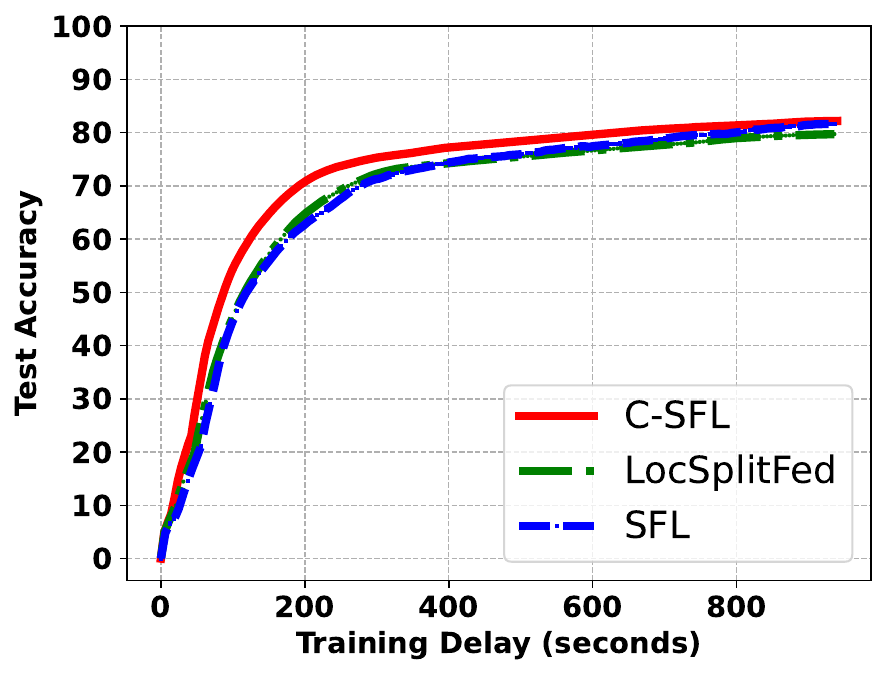} \\
\hspace{-0.05cm} \footnotesize (a) ($\gamma=1$, R=2)   & \hspace{-1.cm} \footnotesize (b)  ($\gamma=8.5$, R=10) & \hspace{-1.cm} \footnotesize (c) ($\gamma=1$, R=10)\\
\end{tabular}
\caption{Effect of client heterogeneity ratio ($\gamma$) and transmission rate $R$ during the training on the FMNIST. C-SFL (Ours) is mostly beneficial when the client heterogeneity is high and the transmission rate is small (e.g. mobile/IoT devices).}
\label{fig:exp2}
\vspace{-0.3cm}
\end{figure}

\textbf{Effect of client heterogeneity $\gamma$ and transmission rate $R$:} In Fig.\ref{fig:exp2} we show the performance of each scheme as a function of the client heterogeneity ratio $\gamma$ and the transmission rate $R$.
The experimentation is based on two extreme cases: 1) high heterogeneity $(\gamma=8.5)$ and 2) no heterogeneity $(\gamma=1)$, meaning all clients have the same computing speed ($p_n=16 Ghz$).
We observe that in ideal cases with clients with the same computing speed ($\gamma = 1$) and high transmission rates ($R=10$), the accuracy gains of our scheme are small, compared to the other two schemes (see Fig. \ref{fig:exp2}c). 
The values of $h$ and $v$ are shown in Table \ref{tab:cut_layers} for each scheme and case. 
However, the accuracy gains of our scheme are more in cases with high or no heterogeneity among clients ($\gamma =8.5$) and low transmission rates, as shown in Fig. \ref{fig:exp1delay}b and Fig. \ref{fig:exp2}a respectively. 
Interestingly the resulting values of $h$ and $v$ of our scheme change from $(h,v) = (3,4)$ to $(3,5)$, as shown in Table \ref{tab:cut_layers}, when $\gamma$ and $R$ decreased.
Consequently, the aggregator-side model is expanded to more layers, compared to cases without heterogeneity and high transmission rates.
The overall results confirm the advantage of our scheme in several cases, especially in cases with high client heterogeneity and low transmission rates.
\section{Conclusion}\label{sec:conc}
We proposed a new Split Federated Learning scheme that accelerates the training process, reduces the communication overhead and improves model's accuracy.
The key is idea is rather simple: we split the model into three parts, namely the model parts executed at the weak clients, the ones executed at the aggregators, and the ones at the server. Then, we train and aggregate in parallel the aggregator-side and server-side models advocating a local loss function. Through experiments, we validated the superiority of our method in terms of training delay and communication overhead against the existing SFL methods.

We have just scratched the surface of designing a more composite hierarchical architecture for Split Federated Learning.
A natural extension is to design an algorithm for finding the optimal cut and collaborative layers for the case that the former are the same across weak clients, and the latter are the same across the local aggregators as in the case in this paper. The next step would be to find different layers for different devices.
Another extension is the optimization of the local aggregators' placement and the respective client-to-aggregator assignment.
\subsubsection{Acknowledgements:} This work was conducted in the context of the Horizon Europe project PRE-ACT (Prediction of Radiotherapy side effects using explainable AI for patient communication and treatment modification). It was supported by the European Commission through the Horizon Europe Program (Grant Agreement number 101057746), by the Swiss State Secretariat for Education, Research and Innovation (SERI) under contract number 22 00058, and by the UK government (Innovate UK application number 10061955).
\bibliographystyle{splncs04}
\bibliography{references} 

\begin{thebibliography}{10}
\providecommand{\url}[1]{\texttt{#1}}
\providecommand{\urlprefix}{URL }
\providecommand{\doi}[1]{https://doi.org/#1}

\bibitem{collins:2022}
Collins, L., Hassani, H., Mokhtari, A., Shakkottai, S.: Fedavg with fine tuning: Local updates lead to representation learning. Advances in Neural Information Processing Systems  \textbf{35},  10572--10586 (2022)

\bibitem{fu:2023}
Fu, L., Zhang, H., Gao, G., Zhang, M.: Client selection in federated learning: Principles, challenges, and opportunities. IEEE Internet of Things Journal  (2023)

\bibitem{han:2021}
Han, D.J., Bhatti, H.I., Lee, J., Moon, J.: Accelerating federated learning with split learning on locally generated losses. In: ICML 2021 workshop on federated learning for user privacy and data confidentiality. ICML Board (2021)

\bibitem{imteaj:2021}
Imteaj, A., Thakker, U., Wang, S., Li, J., Amini, M.H.: A survey on federated learning for resource-constrained iot devices. IEEE Internet of Things Journal  \textbf{9}(1),  1--24 (2021)

\bibitem{jiang:2022}
Jiang, Y., Wang, S., Valls, V., Ko, B.J., Lee, W.H., Leung, K.K., Tassiulas, L.: Model pruning enables efficient federated learning on edge devices. IEEE Transactions on Neural Networks and Learning Systems  \textbf{34}(12),  10374--10386 (2022)

\bibitem{kim:2023}
Kim, M., DeRieux, A., Saad, W.: A bargaining game for personalized, energy efficient split learning over wireless networks. In: 2023 IEEE Wireless Communications and Networking Conference (WCNC). pp.~1--6. IEEE (2023)

\bibitem{lin:2024}
Lin, Z., Qu, G., Wei, W., Chen, Cheng, X.: Adaptsfl: Adaptive split federated learning in resource-constrained edge networks. arXiv preprint arXiv:2403.13101  (2024)

\bibitem{lin:2024efficient}
Lin, Z., Zhu, G., Deng, Y., Chen, X., Gao, Y., Huang, K., Fang, Y.: Efficient parallel split learning over resource-constrained wireless edge networks. IEEE Transactions on Mobile Computing  (2024)

\bibitem{mcmahan:2017}
McMahan, B., Moore, E., Ramage, D., Hampson, S., y~Arcas, B.A.: Communication-efficient learning of deep networks from decentralized data. In: Artificial intelligence and statistics. pp. 1273--1282. PMLR (2017)

\bibitem{nokland:2019}
N{\o}kland, A., Eidnes, L.H.: Training neural networks with local error signals. In: International conference on machine learning. pp. 4839--4850. PMLR (2019)

\bibitem{Seungeun:2022}
Oh, S., Park, J., Vepakomma, P., Baek, S., Raskar, R., Bennis, M., Kim, S.L.: Locfedmix-sl: Localize, federate, and mix for improved scalability, convergence, and latency in split learning. In: Proceedings of the ACM Web Conference 2022. pp. 3347--3357 (2022)

\bibitem{one6g:2024}
one6G, w.p.: 6g technology overview  (2024)

\bibitem{samikwa:2022}
Samikwa, E., Di~Maio, A., Braun, T.: Ares: Adaptive resource-aware split learning for internet of things. Computer Networks  \textbf{218},  109380 (2022)

\bibitem{shin:2023}
Shin, J., Ahn, J., Kang, H., Kang, J.: Fedsplitx: Federated split learning for computationally-constrained heterogeneous clients. arXiv preprint arXiv:2310.14579  (2023)

\bibitem{thapa:2022}
Thapa, C., Arachchige, P.C.M., Camtepe, S., Sun, L.: Splitfed: When federated learning meets split learning. In: Proceedings of the AAAI conference on artificial intelligence. vol.~36, pp. 8485--8493 (2022)

\bibitem{tirana:2024mp}
Tirana, J., Lalis, S., Chatzopoulos, D.: Mp-sl: Multihop parallel split learning. arXiv preprint arXiv:2402.00208  (2024)

\bibitem{tirana:2024}
Tirana, J., Tsigkari, D., Iosifidis, G., Chatzopoulos, D.: Workflow optimization for parallel split learning. IEEE INFOCOM 2024-IEEE Conference on Computer Communications pp. 1331--1340 (2024)

\bibitem{vepakomma:2018}
Vepakomma, P., Gupta, O., Swedish, T., Raskar, R.: Split learning for health: Distributed deep learning without sharing raw patient data. arXiv preprint arXiv:1812.00564  (2018)

\bibitem{wang:2021}
Wang, S., Zhang, X., Uchiyama, H., Matsuda, H.: Hivemind: Towards cellular native machine learning model splitting. IEEE Journal on Selected Areas in Communications  \textbf{40}(2),  626--640 (2021)

\bibitem{wang:2023}
Wang, Z., Xu, H., Xu, Y., Jiang, Z., Liu, J.: Coopfl: Accelerating federated learning with dnn partitioning and offloading in heterogeneous edge computing. Computer Networks  \textbf{220},  109490 (2023)

\bibitem{wu:2023}
Wu, W., Li, M., Qu, K., Zhou, C., Shen, X., Zhuang, W., Li, X., Shi, W.: Split learning over wireless networks: Parallel design and resource management. IEEE Journal on Selected Areas in Communications  \textbf{41}(4),  1051--1066 (2023)

\bibitem{xu:2023}
Xu, C., Qu, Y., Xiang, Y., Gao, L.: Asynchronous federated learning on heterogeneous devices: A survey. Computer Science Review  \textbf{50},  100595 (2023)

\bibitem{ye:2023}
Ye, M., Fang, X., Du, B., Yuen, Dacheng: Heterogeneous federated learning: State-of-the-art and research challenges. ACM Computing Surveys  \textbf{56}(3),  1--44 (2023)

\bibitem{zhang:2021}
Zhang, C., Xie, Y., Bai, H., Yu, B., Li, W., Gao, Y.: A survey on federated learning. Knowledge-Based Systems  \textbf{216},  106775 (2021)

\end{thebibliography}
\end{document}